# Don’t Fake It If You Can’t Make It: Driver Misconduct in Last-Mile Delivery

Srishti Arora,[a] Vivek Choudhary,[b] Pavel Kireyev[a]
[a]INSEAD, Singapore; [a]NBS, NTU, Singapore

***Abstract***

In the last two decades, last-mile delivery (LMD) firms have seen immense growth fueled by the success of e-commerce, leading to faster and cheaper deliveries. Operating on thin margins, LMD firms strive for successful first-time deliveries to avoid the financial and reputational costs of reattempts. Delivery Agents (DAs) are integral to LMD efficiency, influencing customer experience, delivery success, and productivity. However, most LMD performance enhancement research focuses on process, technology, and incentives, which presume workers will conform to procedures and monitoring tools will function flawlessly. Nevertheless, in practice, DAs deviate from expected behaviors, i.e., indulge in misconduct, negatively affecting delivery efficiency, often resulting in returned parcels. One of the major misconducts is fake remarked deliveries, wherein DAs intentionally do not deliver the parcels and provide a fake reason for it. For instance, even without reaching a delivery address, a DA remarks ‘customer unavailable’ and records a delivery failure. In this study, we collaborated with a leading Indian LMD firm and, using instrumental variable regression, find that such misconduct leads to a spillover productivity loss. This effect reduces the next day’s successful deliveries by 1.60% and first-time-right deliveries by 1.86%. We discuss misconduct’s correlation with factors such as task complexity and offer novel insights into how opportunistic circumstances can influence worker behavior.

---

*Keywords: Last-mile Delivery, Platforms, Behavioral Operations, Misconduct, Productivity*

## 1. Introduction

Global e-commerce is projected to grow 8% annually from 2020 to 2025, triggering a 36% increase in delivery vehicles across the world’s 100 largest cities by 2030 (WEF 2020). Last-mile delivery (LMD) is the most complex and expensive part of the e-commerce supply chain (Seghezzi and Mangiaracina 2022) and holds practical and academic significance. It presents strategic value for businesses seeking competitive advantage (Harrington 2019) despite challenges like high fragmentation (Winkenbach 2019) and limited observability.[1] Most e-commerce companies partner with LMD firms to manage deliveries. These LMD firms encounter numerous challenges, including competition, slim profit margins, and increasing customer expectations, despite a lack of willingness to pay for these enhancements (Mangiaracina et al. 2019).[2] As a result, cost-effective strategies to maximize LMD efficiency are highly desirable.

[1] https://www.supplychainbrain.com/blogs/1-think-tank/post/32800-last-mile-delivery-challenges-and-how-to-solve-them

[2] https://www.forbes.com/sites/forbesbusinessdevelopmentcouncil/2021/05/06/massive-growth-challenges-and-opportunities-for-third-party-logistics-post-pandemic

For delivery, LMD firms employ contractual gig workers (Altenried 2019), often called delivery agents (DAs). DAs are instrumental to LMD efficiency, and their conduct directly impacts the company's reputation and profitability. The existing efforts to enhance LMD performance, such as route optimization and alternative delivery methods (Olsson et al. 2019), assume a precise adherence to procedures by DAs. Yet, the realities of the gig economy suggest that DAs often fail to adhere to established processes. As a result, LMD firms identify misconduct as one of the primary reasons for inefficient delivery and the subsequent return of parcels to merchants. However, the impact of DA misconduct - a pervasive issue in the gig economy (Burbano and Chiles 2022) - remains understudied. Consequently, this lack of research limits our understanding of its implications for delivery efficiency. Our study aims to bridge this knowledge gap by focusing on misconduct, which has substantial operational implications (Chan et al. 2021).

Misconduct can inflate operating costs, increase carbon footprint, provoke customer churn, and tarnish the relationship with e-commerce partners. However, in many contexts, firms cannot use penalties to deter such behavior, as the DAs can easily switch to other platforms due to the contractual nature of the job and abundant job opportunities. Furthermore, the negative implications of increased monitoring of gig workers (Liang et al. 2022) may outweigh the benefits. This is the focal point of our paper: to quantify the impact of misconduct on LMD performance and identify conditions that can exacerbate or mitigate this impact.

In collaboration with a leading LMD firm in India, we study DA misconduct, specifically "fake remarks," a crucial contributor to parcel delivery failures and eventual returns to merchants. For each unsuccessful attempt, a DA records a remark indicating failure reasons. DAs can be honest, stating 'could not deliver,' or dishonest, making a false remark like 'the customer was not available' without even visiting the customer's address. LMD firms allow three delivery attempts before returning a parcel to the e-commerce platform to account for genuine failures. These returns incur significant costs, affecting revenues, reverse logistics, and the company's reputation. However, this allowance unintentionally encourages fake remarks, which we frequently observe in our data. This issue of fake remarks is increasingly acknowledged in the industry, and several startups[3,4] are attempting to address it through technical solutions or manual verification methods. However, infrastructural limitations in India, the large volume of parcels, inaccuracies in postal codes, high GPS error rates, and unstructured addresses undermine the effectiveness of these solutions, thereby providing an easy loophole for DAs to evade accountability. But this issue isn't restricted to India; there are similar complaints about missing parcels and incorrect delivery updates globally.[5,6] Our paper delves into the impact of fake remarks on firm performance and seeks to understand the motivations driving DAs to engage in this misconduct.

[3] https://www.clickpost.ai/blog/top-8-reasons-for-ndr-in-ecommerce-and-how-to-solve-them#title_11
[4] https://www.shiprocket.in/blog/know-how-you-can-prevent-fake-delivery-attempts/

We employ an instrumental variable regression approach to identify the spillover effect of fake remarks on different delivery outcomes, such as overall delivery success and first-time-right success. Our results show that fake remarked deliveries have a cascading effect, decreasing overall successful deliveries by 1.6% on the following day. This effect is explained by effort reallocation to conceal misconduct. This misconduct spills over, causing a 1.86% decrease in first-time-right (FTR – successful deliveries in the first attempt) deliveries the next day, significantly affecting firm efficiency given the FTR's role in supply chain efficiency (Rautela et al. 2021).[7] This seemingly small percentage can cost millions of dollars annually. We explore how workers adjust their behavior to their peers and how opportunistic circumstances such as cash-on-delivery parcels can induce misconduct, exacerbating the impact of fake remarks. We also discuss how misconduct is associated with factors such as task complexity, loyalty, and professionalism to provide actionable insights. These findings suggest the need for strategic allocation of parcels, prioritizing FTR success, and recognizing situations favorable to misconduct when designing processes. Counterintuitively, the effect of fake remarks worsens when the DA is familiar with the area, suggesting DAs with misconduct history should be allocated to newer areas. Our work provides insights into the impact of aberrant behaviors on workers' productivity in last-mile logistics, offering valuable perspectives for academia and industry. To the best of our knowledge, we are the first to analyze, quantify, and provide circumstances that exacerbate misconduct and its impact on LMD productivity, notably affecting operational costs.

## 2. Last-Mile Delivery Process

Our collaborator provides logistics for major e-commerce entities. As of June 2021, its expansive network includes over 2,000 delivery centers across 15,000+ postal codes in tier 1 and 2 cities. When a consumer orders on the partner e-commerce website, the LMD firm collects the parcel from the designated warehouse, routing it to the delivery center via regional centers. The delivery center is the last stop for the parcel before it gets delivered to the customer by the DAs (i.e., the last-mile leg, which is the focus of this paper). The LMD firm employs contractual DAs, who primarily use two-wheelers for deliveries.

Center managers, upon recruiting a DA, allocate them to an exclusive area (non-overlapping between DAs) within their center's coverage region, reflecting demand needs and the DA's familiarity with the area. The intent is to balance parcel distribution amongst all the DAs by apportioning varied geographical territories. Such division of exclusive areas ensures a nearly equal workload across DAs.[8] Once contracted,

[5] https://www.yugatech.com/business/why-your-lazada-orders-are-flagged-as-delivered-even-if-theyre-not/
[6] https://thepackageguard.com/amazon/amazons-delivery-service-can-sometimes-leave-guessing/
[7] https://postandparcel.info/93399/news/e-commerce/true-cost-implications-failed-deliveries/
[8] Meaning each DA gets 8-hours of workload every day.

DAs must report to their delivery centers every morning when they pick up the parcels, and in the evening, when they return the undelivered parcels. The manager overseeing the center assigns each day's dispatch, which is a collection of parcels, to the DA. The allocation of dispatches takes no account of any backlogs; every delivery designated for a particular area on a given day is entrusted to the DA. On average, a mere 8.4% of the parcels are carryovers, with the majority being fresh parcels. To discourage intentional carryover, fresh parcels are supplied regardless of the presence of any carryovers. DAs have no control over the quantity of parcels, as these arrive exogenously based on customer demand in the area. This is independent of the work allocation and the behavior of the DAs.

As illustrated in Figure 1, a DA handles three parcel types daily – a) fresh and unattempted parcels to be attempted for the first time, b) fake remarked reattempt parcels, previously fake remarked by the same DA, and c) genuine reattempt parcels, undelivered previously by the same DA because of genuine reasons. An undelivered parcel is re-assigned to the same DA the following day. After the allocation, DAs have the autonomy to determine the parcel delivery sequence and route. Naturally, while many DAs prioritize proximity to the delivery center, others may consider parcel type (e.g., carryover first followed by fresh). Our DA interviews corroborate these strategies.[9] Despite this autonomy, the firm requires that each parcel is attempted every day and that any issues are reported transparently. If DAs are unable to attempt a parcel due to certain constraints, they can leave the remark "*not attempted*."

Although "*not attempted*" remarks are permitted, they are seldom used due to the potential for managerial reprimand and lack of incentive. According to our interviews with DAs,[10] in practice, all parcels are expected to be attempted, regardless of the hours worked. Section 3.1 explores this, corroborating the absence of such entries. Therefore, "*not attempted*" was not included in our analysis. However, some DAs leave parcels untouched without taking any action; we refer to such parcels as "*unattempted*." Note that not every parcel delivery attempt culminates in success. Parcels from unsuccessful delivery attempts are returned by the DA to the delivery center at the end of the day for subsequent tries on the following days. After three unsuccessful attempts, parcels are returned to the merchant.[11]

To facilitate the delivery process, the firm provides a smartphone app to the DAs. The DAs can view each parcel's attempt history in their app. Figure A.1a in the online appendix[12] shows the screenshots of the DA app displaying parcels and delivery details. DAs' can record delivery status and contact the consignees via the app. DAs are asked to call customers before attempting delivery. Calls can be made through the app, and the firm tracks their duration. When a DA attempts a delivery, they need to enter a remark indicating success or failure (Figure A.1b). For unsuccessful deliveries, they must select a reason from a pre-populated list of remarks within the app (Figure A.1c), such as "Consignee unavailable" and "Entry restricted area."

---

[9] "I deliver block by block based on distance from the delivery center" and "I deliver first in the nearest areas but attempt the carryover parcels before the fresh if the parcels are from the same area."

[10] "I have no choice but to deliver all parcels assigned to my area," and "Regardless of the load, we have to attempt all the parcels."

[11] This is a common agreement and a norm between e-commerce firms and LMD firms.

[12] Note: All tables, figures, and sections having a prefix "A." are available in online appendix (e-companion).

**Figure 1: Dispatch Lifecycle**

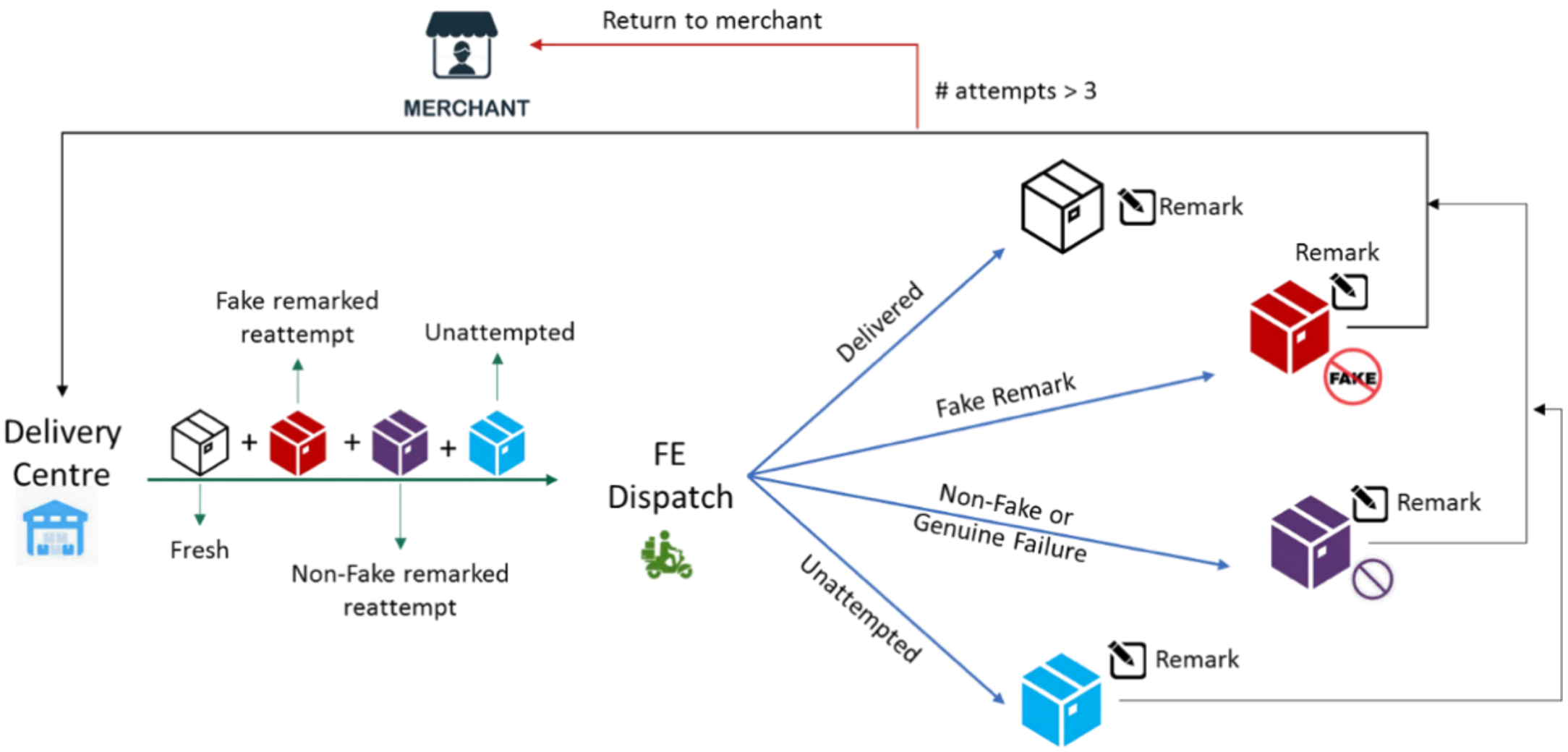

According to the firm, the largest portion of the parcels returned to the merchants is due to fake remarks, a pervasive issue confronting all industry players. Consequently, the firm conducts a daily audit of failed deliveries and seller complaints of false attempts. Yet, due to the massive scale of operations, full auditing is impractical and limited to the center manager making random calls to consignees to verify failure reasons. Moreover, full auditing is avoided to contain costs and prevent tarnishing customer perceptions, acknowledging that transparency doesn't always ensure satisfaction (Bray 2020). Excessive auditing might even provoke a perception of misconduct.[13] After years of refinement, the firm has developed an algorithm to identify remarks as "fake," classifying them based on the DA's location during the remark entry and call records to the consignee. Table A.1 illustrates the four conditions for identifying a fake remark. As fake remark audits are random, DAs cannot predict which deliveries may get audited, and the fake remark detection logic is not disclosed to them to avoid gaming. Only the daily identified fake remarks are communicated to the DAs, and weekly performance reports are generated for them and their delivery center.

[13] Decrinis (2022) explains that enforcement systems, such as manual auditing, that rely on rewards and penalties, might aid ethical fading, as workers might feel controlled and thus resort to more misconduct under high behavioral invisibility. As we discuss in section 3, the choice to fake a delivery depends on the behavioral invisibility of the parcel. Bernstein (2012) echoes this, stating that more observability is sometimes detrimental to worker performance, as it induces strategic behavior due to perceived control. Liang et al. (2022) reported that the downside of increased monitoring in a gig economy makes it infeasible to keep a check on workers' behavior, and Burbano and Chiles (2022) explained that the threat of monitoring reduces employees' trust of employers.

DAs receive a fixed salary with a minor variable component tied to successful deliveries. This variable component is inconsequential compared to their fixed pay, as supported by DA interviews.[14] The focal firm maintains this pay structure due to hiring challenges; initial attempts at a large variable component led to inconsistent compensation and driver dissatisfaction.[15] Further, Eisenhardt (1989) provides a compelling theoretical argument against a variable pay structure for LMD. While outcome-based contracts align the interests of the agent with those of the principal by transferring risk to the agent, they motivate only when outcome uncertainty is low. As uncertainty increases, the cost of transferring risk outweighs these contracts' motivational advantages. In parcel delivery, outcome uncertainty is high due to fluctuating volumes and external factors such as customer availability, road closures, etc. Therefore, the costs of risk transfer to agents might result in unintended consequences, as discussed above, such as competition for delivery areas, pay disparity, and financial instability amongst DAs. Given the minimal impact of misconduct on incentives in our setup, we can distinctly isolate and understand the effect of misconduct on performance.

## 3. Literature and Hypotheses Development

Our work intersects with three research areas: last-mile delivery, people-centric operations, and unethical behavior. Management scholars explore how LMD parameters, like operational transparency (Bray 2020), delivery speed (Fisher et al. 2019), promised delivery date (Salari et al. 2022), and on-time delivery (Liu et al. 2021), can improve performance metrics such as sales. Moreover, research on delivery worker behavior within LMD is growing. It has identified the effects of factors such as geographical familiarity (Mao et al. 2019) and ratings and penalties (Xu et al. 2020) on LMD performance. A few papers address operational failures in LMD, such as the implications of failed delivery attempts (Lim et al. 2023) and order fulfillment delays on customers' anxiety (Rao et al. 2011). Yet, to our knowledge, no study has examined the impact of workers' misconduct and strategic behavior on operational efficiency in LMD.

Additionally, LMD's environmental sustainability and topics like traffic congestion and emissions (Mucowska 2021) have been explored. Proposed solutions include customer pickup lockers (Lyu and Teo 2022), electric vehicles (Siragusa et al. 2022), and sharing environmental impact information with customers (Ignat and Chankov 2020). FTR failures, estimated between 10% and 50% (Voigt et al. 2022), not only increase operating costs (Hasija et al. 2020) and customer attrition but also considerably increase emissions (Edwards et al. 2009). However, existing studies overlook emissions from failed deliveries and the role of DA misconduct in such failures.

[14] "We get 7 Rs (0.08 USD) per parcel after the target. My incentive was just 184 Rs (2.27 USD) for this month. It is insignificant for me. Target is usually too high to earn any additional incentive (beyond my fixed salary)" and "Irrespective of the load we have to attempt all the parcels, so why think about incentive."

[15] As LMD firms compete for workers within India's growing economy, the firm aims to mirror the stability of a fixed-salary job, offering monthly contracts over per-piece remuneration. Given the defined delivery areas but uncertain volume, some DAs might end up getting higher enumerations than others in piece rate compensation. This could spark competition for delivery areas and favoritism by the center managers. By minimizing variable pay, the firm prevents competition for delivery areas, pay disparity, and financial instability. This policy also encourages the attempt of all parcels, eliminating daily income targeting. Salaries are consistent across all centers but adjusted to city's cost-of-living index. There was no change in incentive structure during our study, though the firm withheld incentive data due to confidentiality.

The second stream of literature to which our work contributes is People Centric Operations (PCO), defined by Roels and Staats (2021) as "the study of how people affect the performance of operational processes." Most PCO studies focus on decision-making processes to improve performance. However, workers' strategic behavior and its implications on firm performance have been studied sparingly in the OM literature (Roels and Staats 2021). For instance, a few studies on employee strategic behavior examine how an increased workload's impact on performance can be explained by a physician's task completion preferences (KC et al. 2020) and how worker misconduct varies in the presence of counterproductive peers (Chan et al. 2021). Our paper addresses these gaps by investigating the impact of strategic behavior on operational efficiency and demonstrating its adverse effect on a firm's performance.

Our work also intersects with a third stream of literature focusing on unethical or aberrant behavior. Aberrant behavior[16] leading to decreased worker productivity has been conceptualized in various ways across economics and business literature (Eliyana and Sridadi 2020). The literature also provides the theory on why workers indulge in misconduct. For instance, Pierce and Balasubramanian's (2015) review suggests that social and professional interactions can increase misconduct within organizations, with social comparison serving as an additional catalyst. Following Kidwell and Bennett (1993), we connect the theory of unethical behavior to LMD, where worker behavior is crucial. They postulate several causes for misconduct: first, the focal worker might withhold effort if they believe others are doing so; second, they might limit effort when they perceive the task to be less observable to the supervisor; and third, the worker's effort can be positively influenced by the perceived lack of alternative employment opportunities. Overall, the OM literature has insufficiently examined the impact of workers' aberrant behaviors on their productivity, which we aim to contribute to in the LMD context.

### 3.1. Theoretical Framework and Hypotheses Development

The LMD setting is typically characterized by granular audits and a plethora of alternative employment opportunities, potentially fueling a tendency to withhold effort i.e., shirk. This paper focuses on an interesting and quantifiable instance of shirking: "fake remarks" input by the DAs.

A DA might input a fake remark for various reasons, such as unwillingness to travel to a certain delivery area or having to cover a considerable distance for just a handful of parcels. While DAs possess the autonomy to determine the sequence of their delivery attempts, they do not have the discretion to not attempt a parcel, extending their work hours as necessary.[17] We argue that at times, DAs may choose not to attempt certain parcels due to ethical fading, a phenomenon wherein individuals focus on one aspect (such as effort reduction) to the extent that they overlook the ethical implications (like company policy) of their decisions (Tenbrunsel and Messick 2004). While ethical fading has been extensively studied in fields like psychology (Tenbrunsel et al. 2019), economics (Puaschunder 2022), and business ethics (Gunia 2019, Rees et al. 2019), we employ this relevant theoretical construct to shed light on the research question at hand.

[16] Robinson and Bennett (1995) classify behaviors such as excessive breaks or work slowdowns as "production deviance," a subcategory of aberrant behaviors, whereas Kidwell and Bennett (1993) define "shirking" as employees deliberately withholding effort for self-interest and opportunism.

Ethical fading is triggered particularly in complex organizational settings (Decrinis 2022) such as LMD firms. One facet of this complexity is behavioral invisibility (closely related to private information of the agents that leads to information asymmetry), which is defined as the lack of visibility in the actions of the DAs. The firms cannot monitor every action of a DA (e.g., what the customer said to the DA over a phone call). Therefore, parcels that the DA perceives as less likely to generate customer complaints and more likely to be successfully delivered on a subsequent attempt provide the necessary behavioral invisibility, facilitating misconduct. When a DA decides not to attempt a parcel, they have three options: record it as "*not attempted*," take no action (we term these as "unattempted" parcels), or input a "fake remark."

Unattempted parcels represent discretionary non-deliveries. However, when a DA inputs a fake remark, it breaches the code of conduct, and hence, we (and the firm) categorize it as misconduct. Organizational misbehavior encompasses actions that intentionally violate expected conduct (Vardi and Wiener 1996), extending beyond illegal or unethical actions to include deviation from process requirements (Gabbioneta et al. 2019) and misreporting (Amar et al. 2022). Hence, a fake remark is essentially misconduct.

Center managers conduct daily corrective and preventive action (CAPA)[18] meetings, where DAs account for failed deliveries and face reprimand for any misconduct detected during audits or through customer complaints. Conducted in front of co-workers, these discussions induce fear and shame that our interviews with DAs confirmed.[19] Unattempted parcels, visible to the center managers, guarantee reprimand, while fake remarks might not get caught, thereby avoiding immediate reproach. Hence, despite harsher penalties for fake remarks, their uncertain, delayed nature and the aversion to immediate reputational loss might sway DAs towards this choice. This aligns with theories suggesting individuals prefer a higher probabilistic loss over a lower certain loss (Tversky and Kahneman 1992)[20] and may act unethically for short-term personal gains (Gino 2015). Hence, DAs prefer entering a fake remark over leaving the parcel unattempted.

[17] This lack of discretion in deciding whether to attempt a delivery is a standard practice in this industry. Unattempted deliveries typically result in reprimands, as they directly impact service quality and customer satisfaction. This policy underscores the importance of meeting delivery commitments, albeit at the cost of extended workdays.

[18] This is part of the Quality Management System and is an integral part of ISO certification (link).

[19] "At the end of the day, I have to provide explanations to the center manager for all the undelivered parcels in front of everyone. Usually, I work for 10-11 hours because the route is mostly long, and the number of parcels is high. I don't get time to rest. We are expected to attempt all the parcels irrespective of the time spent on duty" and "We get scolded for pending items, so I have to attempt all."

In summary, four key factors influence DAs' decisions to enter fake remarks, as identified from the literature and our DA interviews: a) unwillingness to deliver, b) high probability of next-day success, c) low chances of customer complaints, and d) behavioral invisibility, with the latter being the most critical. However, instead of a fake remark, DAs might leave the parcel unattempted (without any remark) when they are reluctant to deliver, yet other factors, such as low likelihood of next-day success or high risk of customer complaints, are at play. This decision is often strategic; when the likelihood of getting caught is high, leave parcels unattempted to avoid concealing misconduct. Further, fake remarks and unattempted parcels decrease same-day successful deliveries and create backlogs. With the need to conceal misconduct, DAs prioritize fake remarked parcels, strategically reallocating the effort (Larkin and Pierce 2016).

Principal-agent theory is used in operations management to understand strategic interactions (Leider 2018). According to this theory, the agent's (in this case, the DA's) effort to deliver a parcel creates value for the principal (LMD firm), despite the effort being a disutility to the agent (Arrow 1985). Therefore, the DA aims to minimize effort and maximize incentives. Misconduct is a form of agency problem in gig work settings (Burbano and Chiles 2022), often influenced more by social norms and the saliency of misconduct than cost-benefit analysis (Gino et al. 2009). Fake remarks reduce the DA's effort and enhance flexibility while not affecting financial incentives due to a compensation structure involving fixed pay and minor additional pay for successful deliveries. This provides short-term benefits and encourages myopic DA behavior. Given the negligible impact of misconduct on incentives in our setting, we are able to cleanly identify the effect of misconduct on performance irrespective of the incentives.

In our context, we measure the productivity of a DA by the number of successful deliveries, a key productivity metric for any LMD operation. The DA's strategic behavior in effort reallocation is primarily driven by a need to conceal misconduct. Therefore, the DA might make suboptimal decisions regarding her delivery schedule on the following day, allocating extra time and effort towards the fake remarked deliveries. Considering the suboptimal effort allocation, we predict that fake remarks will create a negative spillover effect and adversely affect future productivity. Formally stated:

**Hypothesis 1.** *Fake remarks will have a negative impact on future productivity.*

Our data do not capture many behavioral traits (e.g., the willingness to enter a fake remark) as well as private information of the DAs. Therefore, we conducted interviews with DAs to understand how their behavior is motivated by their experience and information not available to the firm.

---

[20] Loss means reputational loss in our context.

The interview snippets[21] suggest that a DA uses their past experiences with customers and the delivery area to order their deliveries strategically. This also provides grounds for strategically deciding which parcels to fake remark upon, which they will prioritize the following day. Therefore, the DA, who aims to avoid shame and reprimand, is inclined to deliver these parcels diligently. To improve the success of such deliveries, the DA is likely to put extra effort into delivering undetected fake remarked parcels to conceal their misconduct. Ackerman et al. (2020) argue that individual assessment of a task's disutility influences the degree of effort, shaped by factors like personality, interest, and the job's associated punishment/rewards. Similarly, KC et al. (2020) underscore the importance of understanding workers' task completion preferences, particularly under stress or threat conditions, when they have the autonomy to manage their tasks. Given two parcels, one fake remarked and one fresh, the DA will reallocate effort to conceal mistakes, thus exerting more effort in the fake remarked parcels. Therefore, we posit that fake remarks will increase the success of fake remarked reattempt parcels on the subsequent day.

Furthermore, in their attempt to manage backlogs and conceal misconduct, DAs divert effort from fresh parcels, creating a negative spillover from fake remarked parcels. As Larkin and Pierce (2016) explain, workers divert effort from other tasks to cover up misconduct. The increase in fake remarks requires more effort/time from DAs, which we anticipate will negatively impact FTR. We state the hypothesis that unravels the complete narrative of the effort reallocation process, i.e., predicts the effect of fake remarks on the productivity of fake remarked reattempts and FTR deliveries as follows:

**Hypothesis 2.** *An increase in fake remarks will positively impact future fake remarked reattempt success and negatively impact future first-time-right (FTR) delivery success.*

"Opportunism is a behavior motivated by self-interest and takes advantage of relevant knowledge asymmetry to achieve own gains, regardless of the principles," as defined by Luo and Meyer (2017). Williamson (1993) suggests that "economic agents be described as opportunistic, where this contemplates self-interest seeking with guile." Multiple factors can facilitate opportunism. Ping (1993) and Schwartz and Hirschman (1972) correlate opportunism with a low job-switching cost, while Alchian and Demsetz (1972) suggest that information asymmetry or insufficient monitoring facilitates shirking. Therefore, we define opportunistic circumstances as those amplifying knowledge asymmetry and enabling opportunism. In our context, opportunistic circumstances are those where entering a fake remark is easier and the probability of getting caught is lower. This follows the argument of Gino (2015) – that the more opportunities to justify, the more likely an agent will engage in unethical behavior, and Decrinis (2022), who asserts that lower detection risk prompts more misconduct. Thus, we contend that circumstances and environments that increase knowledge asymmetry will facilitate opportunism.

[21] "We have an idea that some customers will not accept the parcel. I have two such parcels today as well – one from c197 and another from d116. These two customers never accept the parcels. But they order frequently. I have to go all the way to their floor, I get tired, and my time is wasted. They don't even pick up the call" and "I know for which customer I can go directly to their address and for whom I need to call 5-10 minutes before I reach there."

Considering DAs as opportunistic agents, we argue that opportunistic circumstances will enhance their opportunism, which will exacerbate (negatively moderate) the impact of fake remarks on productivity losses. Consequently, we hypothesize:

**Hypothesis 3.** *Opportunistic circumstances will negatively moderate the effect of fake remarks on future productivity.*

**Figure 2: Schematic Theoretical Framework**

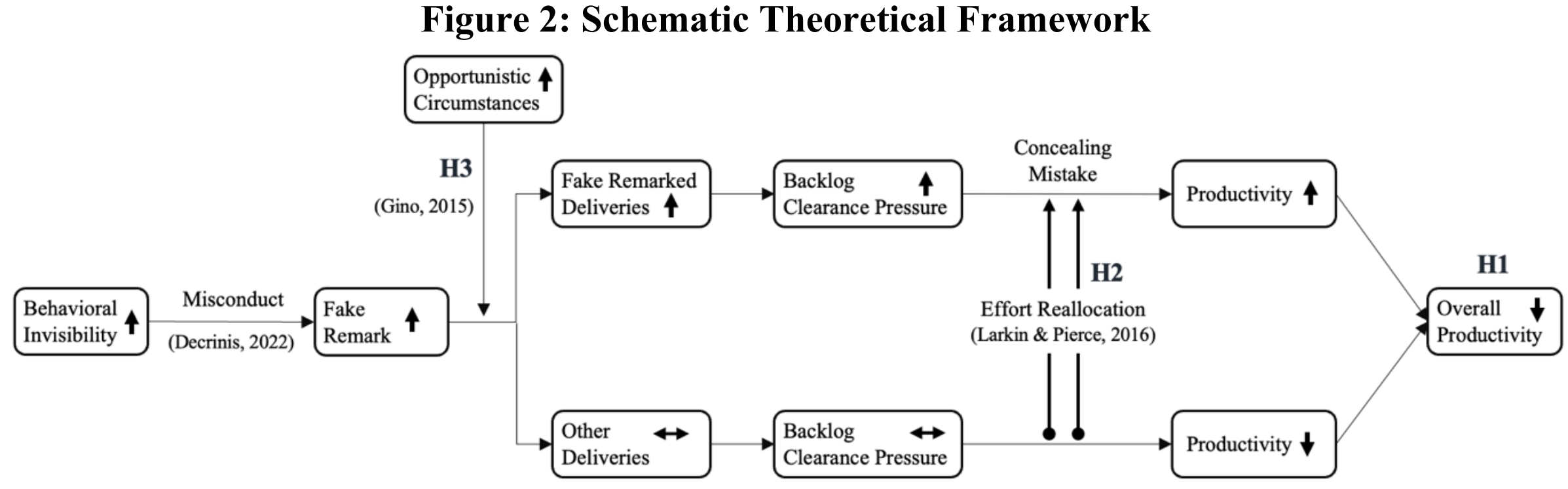

We illustrate the theoretical framework and related hypotheses in Figure 2. It summarizes how behavioral invisibility provides grounds for misconduct, how such misconduct leads to strategic effort reallocation, and how this effort reallocation affects productivity. Finally, it illustrates how opportunistic circumstances may moderate this effect.

## 4. Data and Model

Collaborating with one of the largest LMD firms in India, we obtained our dataset that consists of granular observations for six months (Jun – Nov 2019) from ~900 DAs across nine different cities (both metros and non-metros). During this period, these DAs handled 3.4 million parcels across 71 delivery centers.

Our dataset includes four categories of information: i) *dispatch details* ii) *parcel traces* iii) *DA details,* and iv) *delivery center information*. *Dispatch data* provides a macro view, specifying the assigned DA for a dispatch, parcel counts within a dispatch, and the aggregate number of parcels canceled, delivered, pending, etc. *Parcel traces* offer a micro view, detailing the promised delivery date for each parcel, DA interactions with each consignee (call duration), whether the firm previously served the consignee's address, DA location traces with timestamps for the entire day, and the remark entered by the DA for each parcel. The *DA details* include information such as joining date, work area, and shift timings, while the *delivery center* table provides the dispatch center location and related information.

**Table 1: Descriptive Statistics for Panel Data**

| *Variable* | *Description* | *Min* | *Mean (SD)* | *Max* |
|---|---|---|---|---|
| *success* | number of parcels successfully delivered by the DA in a day, representing the main dependent variable | 6.00 | 38.94 (19.14) | 129.00 |
| *FTR* | number of parcels successfully delivered on the first attempt (first-time-right deliveries) by the DA in a day | 0.00 | 36.67 (18.45) | 127.00 |
| *re_success* | number of parcels that were previously unsuccessful but successfully delivered by the DA in a day, i.e., reattempts that were successfully delivered | 0.00 | 2.23 (2.69) | 40.00 |
| *fr_success* | number of parcels that were previously fake remarked but successfully delivered by the DA in a day. i.e., the total number of fake remarked reattempt successes | 0.00 | 0.31 (0.88) | 24.00 |
| *non_fr_success* | number of parcels that were previously unsuccessful due to genuine reasons but successfully delivered by the DA in a day, i.e., the total number of non-fake remarked successful reattempts | 0.00 | 1.92 (2.36) | 36.00 |
| *genuine_failures* | number of parcels not delivered due to genuine reasons | 1.00 | 19.62 (12.70) | 118 |
| *fake_remark* | number of fake remarks entered by the DA in a day, the primary treatment variable of interest | 0.00 | 2.70 (4.00) | 41.00 |
| *parcels* | number of parcels allocated to the DA in a day | 15.00 | 63.01 (24.50) | 149.00 |
| *da_count* | number of DAs working at the same local delivery center as the focal DA in a day | 1.00 | 6.11 (2.75) | 17.00 |
| *first_attempt* | time (minutes) taken by the DA to make the first delivery attempt of the day, i.e., the time to the first attempt | 2.47 | 59.24 (41.31) | 205.24 |
| *unattempted* | number of parcels not attempted by the DA in a day | 0.00 | 1.75 (2.79) | 34.00 |
| *idle_time* | total idle time (minutes) in a day | 0.00 | 38.4 (27.13) | 228.83 |
| *cod* | number of parcels requiring cash on delivery assigned to the DA in a day | 0.00 | 28.37 (14.67) | 113.00 |

*Note: N=46,086*

In addition to the delivery details, we have obtained data from the DAs' phone GPS traces featuring more than 135 million location traces. This is, to our knowledge, among the most extensive datasets in last-mile logistics research. This helped in calculating key behaviors of DAs such as idle time, time to first attempt, and so on. We also used it to validate fake remarks identification.

### 4.1. Definitions of key variables

We created a day-level panel by aggregating dispatch-level data, where each row corresponds to a DA working on a particular day. Table 1 presents the description of our variables including our dependent variables: *success*, *FTR,* *re_success*, *fr_success*, and *non_fr_success;* main treatment variable: *fake_remark*, and control variables. One should note that *success* is the sum of *FTR* and *re_success,* where *re_success* is the sum of *fr_success* and *non_fr_success*.

We also include several control variables that could potentially affect our dependent variables. Given that workload (KC et al. 2020) and worker availability (Xu et al. 2020) can affect worker behavior, we control for the total number of parcels assigned to the DA and the total number of workers. Additionally, we account for variables such as parcel types – fresh, reattempts, or cash on delivery, the DA's idle time,[22] and the number of unattempted parcels, all of which serve as indicators of the DA's behavior. In sum, our control variables include *parcels*, *da_count*, *first_attempt*, *unattempted*, *idle_time*, and *cod*.

### 4.2. Summary statistics and correlations

Table 1 provides the descriptive statistics for the primary variables. We only consider DAs that have worked on more than 6 dispatches in the six months we observe. The final panel is unbalanced and includes 852 DAs who worked on nearly 2.9 million parcels, yielding 46,086 DA-day observations. On average, a DA works 54 days during our data period, with a mean tenure of 232 days and 25 days off. They often work four-day stretches without breaks and handle about 63 parcels daily, including new and reattempt deliveries. Nearly two-thirds (38.94) of these parcels are delivered successfully, with 94% delivered on the first attempt. We provide a further breakdown of delivery attempt outcomes in Figure A.2.

DAs work an average of 8.15 hours, with the longest recorded workday being 12.15 hours. Each parcel takes an average of 8.8 minutes to deliver. Figure A.3 presents the distribution of daily hours worked by a DA. Note that many addresses order multiple items, enabling faster deliveries. Moreover, the distance between about 43% of parcels in a dispatch is less than 500 meters (an artifact of dedicated assigned areas for each DA). This proximity significantly reduces travel time, supported by the fact that when the number of parcels exceeds 100, the average time per parcel decreases to 4.9 minutes. Figure A.4 highlights that most workloads fall within a manageable range, with the 80th and 90th percentiles at 81 and 97 parcels, respectively. Figure A.5 also demonstrates the remarkable ability of DAs to efficiently manage a substantial volume and size of bags on two-wheelers. Figure A.6 presents examples of small parcels, constituting a substantial portion of the parcels carried by DAs. This greatly facilitates their ability to carry a large number of parcels. This data indicates that although there are days with higher parcel volumes, the distribution and nature of the workload generally remain within a manageable scope for the DAs.

---

[22] Idle time is defined as the time during which a DA does not maneuver on her vehicle between two deliveries. The firm allows up to 15 minutes time per delivery for reasons like walking between apartments in the same building, waiting at the door, and so on, after the DA has reached the location. In all other cases, idle time is calculated using GPS traces. To do so, the duration for which the DA's vehicle moves slower than a threshold speed of 6 km/hr (Pons et al. 2019) is classified as idle time. The firm calculates idle time as the proportion of a rolling 5-minute time window in which the DA is moving at a speed < 6km/hr based on the distance traveled by the DA.

$$idle_time = \begin{cases} \left(1 - \left(\frac{speed\ of\ FE\ in\ km/hr}{6}\right)\right) * (time\ window), & if\ speed < 6km/hr \\ 0, & otherwise \end{cases}$$

We observe an average of 2.7 fake remarks per DA daily, although this number can occasionally reach 41. During our study period, 98.3% of the DAs enter at least one fake remark, and a total of 1.77% of parcels are fake remarked. DAs operate in pre-defined work areas, receiving parcels specific to that area. On average, 88% of the assigned parcels belong to previously visited sub-areas. Also, 45% of the parcels required the DA to collect cash on delivery. Pairwise correlations reported in Table A.2 reveal no issues of multi-collinearity in our data. In Figure A.7, we provide the distributions of all variables.

### 4.3. Econometric Model & Identification

#### 4.3.1. Econometric Model

We identify the impact of fake remarked parcels on the next day's productivity of the DA using a fixed-effect linear regression model:

$$DV_{it} = \beta_1 fake_remark_{i(t-1)} + \beta_2 controls_{it} + fe_i + date_t + c_i w_t + \varepsilon_{it} \quad (1)$$

In equation (1), index $i$ represents a DA, and $t$ denotes a day. $DV_{it}$ represents our dependent variables: *success, FTR, fr_success,* and *non_fr_success*. Using the DV *success,* we estimate the impact of fake remarks on the next day's delivery success, while the other three DVs provide the effect on specific components of success.

$fake_remarks_{i(t-1)}$, our treatment variable, is lagged by one day for each DA, allowing us to estimate the impact of the previous day's fake remarks on today's productivity. As described in section 4.1, we incorporate several controls that could affect the delivery success, specifically *parcels, da_count, first_attempt,* and *cod.* In addition to these contemporaneous variables, we control for lagged variables, *lag.unattempted* and *lag.idle_time,* that could impact the next day's success. We also control for time-invariant and time-variant fixed effects, i.e., DA ($fe_i$), date ($date_t$), and center-week ($c_i w_t$) fixed effects.

#### 4.3.2. Endogeneity

The decision to enter a fake remark is likely endogenous because of the archival nature of our data. Several threats to the identification of $\beta_1$ exist, including sample selection, unobserved heterogeneity, reverse causality, and omitted correlated variables (Antonakis et al. 2014). These are discussed below, along with our countermeasures.

**Selection Bias –** One may argue there might be an endogenous sorting of DAs across delivery centers. Specifically, DAs likely to enter fake remarks could choose centers in areas where completing a delivery is more challenging. However, this concern is mitigated due to the operating style of our partnered LMD firm. Delivery centers are strategically located within a city, employing DAs familiar with that center's service area. These delivery areas do not overlap, limiting choices for DAs to select a delivery center. We control for DA fixed effects, which absorb time-invariant factors such as the center where the DA is employed.

**Reverse Causality –** While productivity or success is dependent on fake remarks, success can likewise affect a DA's actions. To mitigate this, we use a temporal lag, a common approach in management (Ghose and Han 2011) and economics (Bollinger et al. 2020) research. We study the impact of the previous day's fake remarks on the next day's productivity. Moreover, DAs are not privy to tomorrow's parcel assignments, precluding them from entering fake remarks in anticipation of tomorrow's success.

**Omitted Variable Bias –** Certain DA- and time-varying factors that correlate with both our DV and our treatment variable (*fake_remark)* might be unobservable, posing a risk of inducing omitted variable bias. For instance, "strategically selecting parcels for fake remarks based on behavioral invisibility" is an omitted variable that may impact both the DA's productivity and the likelihood of entering a fake remark. Our interviews[23] and data analysis, discussed in section 5, provide evidence of such strategic behavior.

Figure A.8 illustrates this omitted strategic behavior, which we expect to affect future productivity and propensity to enter fake remarks positively. If this strategic behavior is not accounted for, its positive correlation with subsequent success will be absorbed in $\beta_1$, resulting in an underestimation of the effect of fake remarks on productivity. An alternative source of omitted variable bias may be that DAs tend to enter fake remarks under conditions that might reduce future success (e.g., future traffic congestion in an area). Following a similar reasoning, this might overestimate the impact of fake remarks on productivity, However, our analysis does not support this assertion, as we observe stronger evidence in favor of the strategic behavior mechanism depicted in Figure A.8.

We address these biases using instrumental variable (IV) regression and employing a two-stage least squares (2SLS) regression[24] in the next sub-section. In section 8, we conduct several robustness tests and alternative specifications, such as using propensity score matching, a nonlinear model with a control function, or applying log transformations to our outcome variables, to ensure the consistency of our results.

**Unobserved Heterogeneity –** DAs and delivery centers may be heterogeneous in many ways. To address this, we control for three sets of fixed effects. We use DA[25] and date fixed effects to account for time-invariant heterogeneity. Following Bai (2009), we include center-week fixed effects to capture time-varying unobservables. We opt for week-level interactive fixed effects because weekly reports and audits are in place to analyze and course-correct center performance.

---

[23] "Customers of reputed e-commerce merchants like Amazon are easily available and accept easily. For them, I don't feel the need to make a call. For parcels from other merchants, it is always better to call the customer beforehand."

[24] We use the 2SLS approach based on the recommendations of Angrist and Krueger (2001). They suggest that even if the underlying relationship is non-linear, linear estimation methods like 2SLS can still provide robust estimates for instrumental variable applications, reducing the risks associated with misspecification.

[25] Given that 87% of DAs worked at only one center, the center is largely time-invariant, and hence, DA fixed effects take care of center-level heterogeneity.

**Serial Correlation –** A DA's fake remarks may exhibit serial correlation, or there could be serially correlated unobservables, such as the DA's effort, requiring control measures to identify the impact accurately. Using time variant fixed effects (center-week in our case) can somewhat alleviate this problem (Bollinger and Gillingham 2012), but a better method in dynamic panel models involves incorporating a lagged dependent variable (Angrist and Pischke 2008, Leszczensky and Wolbring 2022). However, this can induce bias (Nickell 1981). Therefore, we adopt Angrist and Pischke's (2008) estimation strategy, which utilizes the bracketing property of fixed effects and lagged dependent variable estimates. This approach is commonly used in empirical studies (Falk et al. 2018). The robustness of our results is verified using this bracketing property in Appendix A1.10.

#### 4.3.3. Instrumental variables

The null hypothesis of the Wu-Hausman test asserts the absence of endogeneity for the variable *fake_remark*. Since we reject the null hypothesis, we identify *fake_remark* as an endogenous variable. We use the fake remarks entered by the co-workers of the focal DA as an instrument for the endogenous variable *fake_remark*. Norton et al. (2003) suggest that peers or co-workers (people we identify with) can stimulate behavioral change. Misconduct and peer effects, while not extensively studied in OM, have been well-studied in various other research streams (Gino et al. 2009, Pierce and Snyder 2008). Complementing this literature, our interviews also suggest substantial peer interaction at the end of the day. As a result, we draw on the peer effects literature to construct our instrument. Similarly, co-workers' incentives (Sinchaisri et al. 2019) and earnings (Xu et al. 2020) have been used as valid instruments in the literature.

We identify co-workers as DAs working in the same delivery center as our focal DA. Given that DAs have different days off during the week and may take occasional leaves, we establish the co-worker set for each DA for each day, including only those who are working that day. Let $C_t(i)$ be the set of co-workers of DA $i$ on day $t$, excluding the DA $i$. The number of co-workers of DA $i$ on day $t$ is denoted by $|C_t(i)|$. Then, we define the variable $co_fr_avg_{it}$ (mean = 2.7, sd = 1.97) as the past seven-day moving average of the fake remarks per co-worker of our focal DA $i$ starting from day $t$, i.e.,

$$co_fr_avg_{it} = \frac{1}{7}\sum_{k=t-6}^{t} \frac{1}{|C_k(i)|}\left(\sum_{j \in C_k(i)} fake_remark_{jk}\right) \tag{2}$$

We provide robustness checks in section 8 using different windows for our moving average such as 1, 3, and 10 days, which yield consistent results.

As shown in Figure A.9, to estimate equation (1) we use $co_fr_avg_{i(t-2)}$ as an instrument for the focal DA's fake remarks on day $t-1$ i.e., $fake_remark_{i(t-1)}$. First, we estimate an OLS model with the endogenous variable $fake_remark_{i(t-1)}$ (fake remarks of the focal DA on day $(t-1)$) as a dependent variable, employing the instrument $co_fr_avg_{i(t-2)}$. From this, we obtain a predicted value of $fake_remark_{i(t-1)}$ as $\widehat{fake_remark}_{i(t-1)}$. Then, using this predicted value, we estimate the effect of fake remarks on DA productivity by employing different dependent variables, such as *success* and *FTR*.

To ensure robust identification, we need to address potential bias inherent in peer effects settings. The main challenges in studying peer effects include reverse causality, correlated unobservables, and selection bias (also known as endogenous group formation) (refer to Bollinger and Gillingham 2012, Bramoullé et al. 2009, Manski 1993). As discussed in the previous section, we address *reverse causality* by using lagged co-worker fake remarks (similar to Bollinger et al. 2020, Ghose and Han 2011), manage *correlated unobservables* with a rich set of fixed effects (similar to Hanushek et al. 2003, Oestreicher-Singer and Sundararajan 2012), and consider *selection bias* less pertinent as DAs cannot choose their center.

A valid instrument must satisfy three conditions for an unbiased estimate. The *exclusion restriction* requires the IV to be uncorrelated with the error term, and the *relevance condition* requires the IV to be correlated with the endogenous variable (Wooldridge 2002). The IV should influence the dependent variable only via the endogenous variable. Lastly, the instrument needs to be sufficiently strong to provide an unbiased estimate of the effect through the instrumental variable regression (Stock and Yogo 2005). Our instrument satisfies these requirements, as confirmed by the first stage of 2SLS; it's relevant (coefficient = -0.402, $p<0.01$) and strong (F-stat = 273.5).

The exclusion restriction stipulates that our IV influences the performance of the focal worker solely through its impact on the worker's fake remarks. We argue that a co-worker's behavior cannot directly influence the future successful deliveries of focal workers because these workers operate in distinct work areas, and the arrival of parcels is random—driven by consumer demand rather than DA decisions (as discussed in section 2). DAs have pre-defined work areas that are assigned an ample workload, thereby preventing interference among the DAs. The distribution of parcels to individual DAs is not dictated by the combination of fresh and carryover parcels; rather, it is primarily determined by the demand specific to each area. While the center manager endeavors to distribute the load of fresh and carryover parcels evenly among the DAs on average, the pre-defined work areas for DAs limit the manager's ability to exercise much discretion. The areas are designed with the expectation of similar workloads, enabling DAs to earn comparable salaries for similar efforts. Our interviews with DAs substantiate this claim. We conduct a thorough analysis of our instrument's relevance, including theoretical assertions and the elimination of spurious correlations in Appendix A1.7. Additionally, we have provided a detailed explanation of the exclusion restriction, including the validation of the instrument ignorability assumption in Appendix A1.8. We also discuss the direction of the peer effect identified in Appendix A1.9. Lastly, to handle serial correlation, we undertake a robustness test with lagged dependent variable estimation, the results of which remain consistent and are discussed in Appendix A1.10.

## 5. Results

We present the estimation results of the impact of fake remarks on productivity.

### 5.1. Effect of Misconduct

We provide the results of our estimation of equation (1) in Table 2. Column 1 shows the results using only the control variables. We observe that an increase in the parcel count assigned to the DA increases successful deliveries (0.439, $p<0.01$). However, the unattempted parcels from the previous day do not significantly affect today's productivity. We also find that an extended duration to attempt the first parcel in the morning correlates with lower productivity (-0.020, $p<0.01$), while a restful or more idle previous day enhances productivity (0.006, $p<0.01$). Cash on delivery positively impacts successful deliveries (0.233, $p<0.01$). We will delve into cash on delivery parcels in section 6. Column 2 includes our treatment variable *lag.fake_remark* and estimates the model using OLS. Our data suggests the previous day's fake remarks do not affect today's productivity. However, this finding suffers from the aforementioned biases.

In column 3, we use 2SLS to estimate equation (1), with the first stage results reported in Table A.8. Our IV, *lag2.co_fr_avg*, is significant and negatively associated with *lag.fake_remark* (-0.402, $p<0.01$). The F-stat 273.5 suggests that our instrument is strong. This confirms the relevance condition of our instrument. The estimate suggests that an increase in average fake remarks by co-workers decreases the focal DA's fake remarks. A plausible explanation is the strategic response of the focal DA, who might fear heightened scrutiny in case of excessive fake remarks at the center. This mirrors Chan et al. (2021) findings in a misconduct context, where workers are less likely to engage in restaurant theft if their co-workers exhibit higher levels of such behavior on a given day, termed 'strategic peer response.' Negative peer effects have also been identified in several studies across psychology and economics (Angrist 2014, Brady et al. 2017, Schreiner and Bremer 2013) with examples from marathon racing (Emerson and Hill 2018) and labor market unethical behavior when under supervision or reputation risk (Pascual-Ezama et al. 2015).

**Table 2: Effect of Fake Remarks on Productivity**

| | *success* | | |
|---|---|---|---|
| | (1) | (2) | (3) |
| *lag.fake_remark* | | 0.021 | -0.625*** |
| | | (0.017) | (0.210) |
| *parcels* | 0.439*** | 0.439*** | 0.439*** |
| | (0.013) | (0.013) | (0.014) |
| *da_count* | 0.070* | 0.070* | 0.074 |
| | (0.042) | (0.042) | (0.047) |
| *first_attempt* | -0.020*** | -0.020*** | -0.022*** |
| | (0.002) | (0.002) | (0.002) |
| *lag.unattempted* | -0.014 | -0.019 | 0.124** |
| | (0.022) | (0.022) | (0.050) |
| *lag.idle_time* | 0.006*** | 0.006** | 0.015*** |
| | (0.002) | (0.002) | (0.003) |
| *cod* | 0.233*** | 0.233*** | 0.247*** |
| | (0.020) | (0.020) | (0.022) |
| *IV* | No | No | Yes |
| *adj-*$R^2$ | 0.823 | 0.823 | 0.807 |
| *N* | 45,121 | 45,121 | 38,473 |

*Note: *** p<0.01, ** p<0.05, * p<0.1; Standard errors are in parentheses and clustered at the DA level. Estimated with DA, date, and center-week fixed effects.*

In the second stage, our estimates suggest that *lag.fake_remark* negatively affects (-0.625, *p<0.01*) productivity,[26] supporting our H1. A unit increase in fake remarks results in a 0.625 decrease in the next day's successful deliveries. On average, a DA is assigned 63.01 deliveries, successfully delivering 38.94. A decrease of 0.625 equals a daily reduction of 1.6% (0.625/38.94) in successful deliveries. During the study, the company was delivering an average of 1.5 million parcels per day. Therefore, an additional fake remark could result in a spillover effect of 24,000 (0.016×1.5 million) fewer successful deliveries each subsequent day, totaling 8.8 million annually. This significant economic impact should be considered in addition to same-day failed deliveries caused by fake remarks.

One could argue that fake remarks are an adaptive strategy required for capacity management. But if they were a capacity management option, an increase in fake remarks could be expected to improve DA's productivity. However, as we show, an increase in fake remarks leads to a 1.6% overall productivity decrease the next day. Additionally, our analysis reveals that the negative impact of fake remarks is similar in high workload conditions (with a statistically-similar coefficient verified using the Welch t-test), as presented in Table A.9. This uniformity across workload levels indicates that fake remarks are not a capacity management strategy. Further, the number of fake remarks is strategically adjusted in response to peers' misconduct, a behavior coherent only if fake remarks are unethical. Moreover, in our main model, we have controlled for workload when estimating the impact of fake remarked deliveries on subsequent success to ensure that workload variations do not influence our findings.

[26]Angrist and Kolesár (2021) explain that using instrumental variable to counteract omitted variable bias occasionally results in a sign and significance flip. Several studies in both economics (Coe et al. 2012, Friedberg 2001) and management literature (Gopalakrishnan et al. 2022, Song et al. 2020, Tan and Netessine 2014) provide examples of econometric analysis where the sign and significance of the endogenous variable morphs from OLS to 2SLS regression analysis. Friedberg (2001) further elucidates how the direction and magnitude of bias yields divergent estimates between OLS and 2SLS estimation. In our case, we anticipate a positive bias given the positive correlation between strategic behavior (i.e., omitted variable) and both the DA's fake remarks and the DA's success.

After estimating the effect of fake remarks on next-day successful deliveries, we want to test if fake remarks lead to a prolonged cascading effect on productivity. To do so, we estimate the effect of the previous days' fake remarks (days t-1 to t-4) on today's (day t) success. These results have been reported in Table A.10 and illustrated in Figure A.11. We find that today's success is affected by fake remarks made over the past three days, with the effect size diminishing progressively until the day (t-4) when it becomes insignificant. Aggregating this effect (0.625+0.611+0.538 = 1.774), we deduce that success on day t is decreased by an average of 1.774 deliveries due to the previous days' fake remarks. Given that the DA is assigned 63 parcels and successfully delivers 39 of them, a decrease of 1.774 corresponds to an average daily decline of 4.5% (1.774 /39) in successful deliveries.

To understand whether DAs' past fake remarks influence their fake remarks and how long this serial correlation influences DAs' behavior, we estimate the correlation between previous days' fake remarks (days t-1 to t-12) and today's (day t) fake remarks. As presented in Figure A.12, we find that today's fake remarks are correlated with fake remarks made over the past 8 days, with the correlation diminishing progressively until the day (t-9) when it becomes insignificant. Although the correlations are statistically significant, their magnitude is relatively small. We also quantify the influence of co-workers' past fake remarks (*lag.co_fr_avg*) on the focal DA's current fake remarks (*fake_remark*), in addition to their own past fake remarks (*lag.fake_remark*). This is because we identify strategic peer behavior in our first stage of 2SLS, which shows that DAs are less likely to engage in misconduct if their co-workers exhibit higher levels of such behavior. As shown in Table A.11, we find that co-workers' past fake remarks negatively influence the focal DA's current fake remarks even more strongly than their own past fake remarks. This finding reaffirms our assertion that fake remarks are indeed strategic unethical behavior of the DAs.

Owing to the significant bias induced by endogeneity, we observe a marked change in the direction (sign) and statistical significance of the endogenous variable as we move from OLS to 2SLS regression analysis. This phenomenon aligns with findings in the literature, including Angrist and Kolesár (2021), Melero et al. (2020), and Tan and Netessine (2014). Further, introducing an instrumental variable could alter the direction and significance of other exogenous variables, especially if the endogenous variable correlates with other variables. Tan and Netessine (2014) have reported similar results. In our scenario, the correlation between *lag.fake_remark* and *lag.unattempted* is 0.22, which explains why we observe a change in the direction and significance of the lag.unattempted coefficient when we move from OLS to 2SLS.

### 5.2. Effect of Misconduct on Different Types of Deliveries

To test H2, we use data concerning parcels that were fake remarked earlier and reattempted today, in addition to the first-time-right parcels. We identify three distinct parcel categories (fresh parcels, fake remarked reattempts and genuine reattempts), effectively decomposing *success* into its three components. We evaluate the effect of fake remarks on these specific delivery types by estimating equation (1) with each of the three outcomes serving as dependent variables. These results are presented in Table 3.

In column 1 of Table 3, *fr_success* has a positive coefficient of 0.120 ($p<0.01$), indicating that parcels previously fake remarked by the DA are more likely to be successfully delivered the following day. Next, we use *FTR* as our dependent variable in equation (1). In column 2, we estimate the effect of an increase in the previous day's fake remarks on today's fresh parcels. Using the same IV, we find that the coefficient is significant and negative (-0.682, $p<0.01$), i.e., an increase in fake remarks decreases *FTR* deliveries by 1.86% (or 0.682/36.67) on the subsequent day. Thus, fake remarked reattempts are positively impacted, and *FTR* deliveries are negatively affected by the increase in fake remarks, corroborating H2.

**Table 3: Effect of Fake Remarks on Different Delivery Types**

| | *Fake remarked success* | *First time right* | *Non-fake remarked success* |
|---|---|---|---|
| | *fr_success* | *FTR* | *non_fr_success* |
| | (1) | (2) | (3) |
| *lag.fake_remark* | 0.120*** | -0.682*** | -0.086* |
| | (0.029) | (0.202) | (0.045) |
| *adj-*$R^2$ | 0.240 | 0.793 | 0.256 |

*Note: *** p<0.01, ** p<0.05, * p<0.1; N=38,473; Standard errors are in parentheses and clustered at the DA level. Estimated with instrumental variable using 2SLS with all controls and DA, date, and center-week fixed effects.*

Column 3 presents the estimate for equation (1) with *non_fr_success* as the dependent variable, which accounts for the delivery success of parcels that genuinely failed earlier. The coefficient is significant at the 10% level, indicating that an increase in fake remarks detrimentally impacts the success of these parcels (-0.086, $p=0.055$). This suggests DAs exert more effort to deliver fake remarked reattempt parcels, thereby negatively affecting their success for other parcels (fresh parcels and genuine reattempts). In summary, among the three parcel types, *fr_success* is positively affected, whereas *FTR* and *non_fr_success* suffer due to an increase in the previous day's fake remarks.

Two possible mechanisms underlie these findings. First, the delivery center manager's pressure to deliver the fake remarked parcels identified in the end-of-day audit might encourage DAs to prioritize them.

Second, DAs might anticipate which of today's parcels are likely to be successfully delivered the subsequent day, thus strategically entering fake remarks for them to alleviate today's workload. We posit that both mechanisms are at play, but we cannot distinctly examine an increase in managerial oversight.

Consequently, we focus on DA strategic behavior, supported by extant literature (discussed in section 3), data patterns (discussed below), and our interviews with DAs.[27] DAs, based on their area experience, identify parcels with higher behavioral invisibility and tend to input fake remarks for them. The greater the probability of success the following day, the lower the risk of discovery and penalty for this misconduct, and hence, a greater likelihood of entering fake remarks. This explains the observed positive impact of fake remarks on the successful delivery of previously fake remarked parcels (*fr_success*) in Table 3.

However, as DAs tactically select fake remarked reattempts, their positive efforts on these parcels result in effort reallocation from fresh parcel deliveries, impacting overall productivity. Hence, the impact of fake remarks on *FTR* deliveries primarily determines a DA's overall productivity loss. DAs are less concerned about the successes of genuine reattempts than FTR deliveries but are more cognizant of the reasons behind these parcels' initial failure. With genuine reattempts, they can often coordinate with customers without additional effort (e.g., a customer may have requested next-day delivery during a genuine attempt). Thus, the information gained somewhat mitigates the negative effect of effort reallocation. This means that there is a risk of wasted effort for genuine reattempts if the parcel remains undelivered, unlike with fresh delivery. As a result, we observe a smaller decline in the success of genuine reattempts compared to FTR. DAs strategically prioritize fake remarked reattempts over genuine reattempts to conceal misconduct and often relegate FTR as their last priority. This approach helps clear the backlog while reducing their likelihood of being caught for fake remarks. It establishes how workers reallocate efforts to conceal fake remarks, a phenomenon that would not occur if fake remarks were just an adaptive strategy and not unethical.

In summary, our findings support our assertion that DAs can anticipate the chances of successful reattempts for fake remarked parcels. Table 3 demonstrates that fake remarks correlate with the DA's knowledge of next-day success probability: fake remarked reattempt success increases while both FTR and genuine reattempt success decreases. Furthermore, as corroborative evidence, we note that the median time required to deliver a fake remarked reattempt parcel is 4.15 minutes, as compared to 5.25 minutes for genuine reattempts and 6.41 minutes for fresh parcels. These statistically distinct values affirm our hypothesis of fake remarks as strategic DA behavior and provide evidence of our mechanism at play.

---

[27] "There are areas where if the customer is not picking the call, the customer will not accept the parcel", "I attempt the carryover parcels because if the customer was not available yesterday, he might not even accept today," "Most of the DAs prioritize carryover parcels to get rid of the pending work" and "I call the next customer just before leaving the previous customer location that I will be coming to deliver your parcel."

### 5.3. Causes of Misconduct and Impact on the Platform

We identified three potential predictor categories that could indicate a tendency to enter fake remarks: loyalty and professionalism, task attributes, and peer behavior. Our findings suggest that DAs with higher loyalty and professionalism – indicators such as longer continuous work stretches and less idle time – are associated with fewer fake remarks. Therefore, the stretch of continuous workdays without breaks, work hours, and idle time during the first month of a DA's tenure can be a few informative indicators to predict future misconduct. Vaughan (1999) associates task characteristics with misconduct, leading us to explore this connection. We observe that greater task complexity and workload variability correspond to increased fake remarks. We investigate the influence of peer behavior on the likelihood of fake remarks. We find that the average number of fake remarks made by co-workers and the proportion of co-workers making fake remarks negatively correlate with the focal DA's fake remarks. This corroborates our mechanism that DAs strategically adjust their behavior to avoid detection, mindful of their peers' conduct. Note that while these are not causal analyses, their associations hold valuable insights for management. Detailed analyses can be found in Tables A.12, A.13, and A.14 of Appendix A2.1.

Next, we investigate how fake remark behavior affects platform performance in the long-term. We label a DA as a "faker" if her fraction of fake remarked parcels exceeds the median. As reported in Table A.15, our finding suggests that despite working the same hours (mandated by the center) and covering the same distance, on average these "fakers" idle more and exhibit lower productivity. In Table A.16, we conduct a sub-sample analysis that indicates that fakers make on average 4.2 fewer successful deliveries than non-fakers, equating to a 10% decrease in productivity (the DA successfully delivers 39 parcels). Thus, DAs with more fake remarks tend to complete fewer deliveries in the long run on the platform. Furthermore, fakers' total work hours are not as productive as those of non-fakers.

Though fake remarked reattempts are not costly in terms of operating expenses, given there is no real attempt made, the number of attempts is inversely related to delivery success. Consequently, there's a higher likelihood that a reattempted parcel may end up being returned to the merchant. This creates an added reverse logistics cost for the firm. Moreover, beyond causing customer dissatisfaction, these fake remarks damage the relationship between the LMD firm and its e-commerce partner, incurring further costs. We also observe a slight negative impact of these fake remarks on the success of genuine reattempts on the subsequent day. Genuine reattempts, in turn, carry significant costs in terms of fuel, labor, and time.

## 6. Effects of Opportunistic Environments

In addition to the main effect, we also examined "opportunistic circumstances" that could exacerbate the effect of fake remarks. These are situations where making fake remarks is easier, and detection risk is low. Although precise measurement of opportunism is challenging, we identified indicators in the parcel delivery context, supported by the literature (Bryson and Forth 2007, Mao et al. 2019) and practice.[28] We specifically analyzed variables v*isited_before* and *cod* for their interaction effects with fake remarks on DA productivity. We mean-centered *cod*, *visited_before*, *lag.fake_remark*, and *lag2.co_fr_avg* to mitigate multicollinearity and ascertain the average effect (Afshartous and Preston 2011, Coulton and Chow 1992).

**Table 4: Effect of Familiarity on Productivity**

| | *success* | |
|---|---|---|
| | (1) | (2) |
| *lag.fake_remark_demean* | -0.510*** | 0.093 |
| | (0.100) | (0.147) |
| *lag.fake_remark_demean* × | | -0.028*** |
| *visited_before_demean* | | (0.004) |
| *visited_before_demean* | 0.866*** | 0.853*** |
| | (0.008) | (0.009) |
| *adj-*$R^2$ | 0.927 | 0.921 |

*Note: *** p<0.01, ** p<0.05, * p<0.1; N=38,473; Standard errors are in parentheses and clustered at the DA level. Estimated with instrumental variable using 2SLS with all controls and DA, date, and center-week fixed effects.*

Each parcel has a consignee address that belongs to a postal code.[29] Our variable v*isited_before* serves as a proxy for the DA's familiarity with the delivery area, representing the total number of postal codes in the dispatch the DA has visited until the previous day. When a DA becomes familiar with the delivery area, it becomes easier for her to shirk, justify lower productivity with plausible excuses, and allocate more effort in concealing prior fake remarks. First-stage estimates reveal that the propensity to enter fake remarks has a positive association with *visited_before* (0.008, $p<0.05$),[30] suggesting that DAs are more likely to enter fake remarks for parcels in a familiar area. In Table 4, we report the results of the second stage. In column 2, we find that *visited_before* negatively moderates the effect of fake remarks (-0.028, $p<0.01$). This suggests that the effect of fake remarks is exacerbated as the DA becomes more acquainted with a delivery area supporting H3. Contrary to the literature suggesting that area familiarity increases delivery efficiency (Mao et al. 2019), our results indicate that if the DA has visited an address before, location familiarity can decrease success (in the presence of fake remarks).

We assess the robustness of our results using COD deliveries in Appendix A2.3. To conclude, the impact of fake remarked deliveries on productivity is exacerbated, and the propensity for fake remarks increases under opportunistic circumstances where it is easy to conceal misconduct. This validates our assertion that fake remarks are indeed unethical behavior. The results support our H3 that productivity loss due to aberrant behavior exacerbates under opportunistic circumstances.

[28] https://tedium.co/2020/06/12/cod-cash-on-delivery-history/

[29] Considering that India's nearly 19,100 postal codes cover approximately 2,973,190 square km, each postal code spans an average of 155 square km (radius of 7 km).

[30] We do not use *visited_before* as a control in our main model due to its high correlation with *parcels* (0.88). Table A.17 provides the first stage of the model with *visited_before* as an additional control.

## 7. The Feasibility of Eliminating Fake Remarks and Counterfactual Outcomes

To conduct a counterfactual analysis on eliminating fake remarks, it is necessary to understand that eliminating fake remarks requires at least one of the three conditions: a) a 100%-accuracy fake remark detection technology, b) all DAs are honest and attempt any delivery with the same genuine effort, or c) the manager fully trusts any unsuccessful delivery remark and unattempted deliveries never lead to reprimand. Achieving 100% accuracy in detecting fake remarks (a) is practically impossible, similar to almost all fraud detection applications (e.g., credit card fraud). Regarding b) and c), expecting complete trust between employer and employee is also not feasible because agents are opportunistic, as argued above.

Thus, eliminating fake remarks in practice is unrealistic. The key takeaway is that fake remarks are a form of unethical behavior and are likely to exist in varying magnitudes. This leads to inefficiencies as DAs attempt to conceal it, affecting the success of other deliveries. We present several pieces of evidence to establish fake remarks are unethical: a) disproportionate impact on different delivery types due to effort reallocation to conceal misconduct, b) persistence under varying workload conditions, c) the impact on productivity is exacerbated, and the propensity increases under opportunistic circumstances where it is easy to conceal misconduct, and d) the number is strategically adjusted in response to peers' misconduct. Thus, inefficiencies caused by concealing misconduct through effort reallocation would not exist had fake remarks been honestly labeled as unattempted or unsuccessful. We expect that in the hypothetical scenario where fake remarks were eliminated, there would be more successful deliveries on the subsequent day.

Therefore, even though eliminating fake remarks is unrealistic, reducing them could significantly improve subsequent day performance by allowing for more honest and efficient effort allocation. This improvement would be a net positive for firms (more honest reporting of actual failures and an increase in following days' successful deliveries), suggesting that any reduction in fake remarks would benefit overall efficiency and performance.

## 8. Robustness Tests

We have conducted several robustness tests to ensure our results are consistent. The first set of analyses provides our identification strategy's robustness. We use two matching methods - nearest neighbor matching (Abadie and Imbens 2011) and coarsened exact matching (Iacus et al. 2012) as alternative identification strategies (Appendix A3.1). To ensure the skewness of the dependent variable does not drive our results, we conducted 2SLS with a logarithmic transformation (Appendix A3.2) (Cachon et al. 2019, Goh et al. 2011). Since our DVs are count variables, we also implemented the control function approach (Wooldridge 2015) using fixed effects OLS in the first stage and negative binomial regression in the second, followed by bootstrap standard errors as suggested by Wooldridge[31] (Appendix A3.2). As robustness to our fake remarks' measurement methodology, we performed sub-sample analysis with a conservative estimate of fake remarks to test for error-free measurement (Appendix A3.3). To ascertain the robustness of our 2SLS result, we varied the length of the time window (1, 3, and 10 instead of 7) over which we calculated the peer effect (co-workers' moving average fake remarks described in equation 2) (Appendix A3.4). Next, we used all the instruments together and conducted the Sargan test, an overidentified exclusion restriction test in the literature (Moon et al. 2022, Wu et al. 2020) (Appendix A3.5).

The next group of analyses is to rule out alternative explanations. We examined whether the promised delivery date and the number of delivery attempts influence the number of successful deliveries or instances of fake remarks (Appendix A3.6). To eliminate concerns of selection bias because of the attrition (Elwert and Segarra 2022), we ran the main model with a sub-sample of DAs who did not quit during our study period, i.e., DAs whose end date was post 30-Nov-2019. Other tests for attrition are also provided in Appendix 3.7. To validate that the days with an unusually high number of fake remarks do not skew our overall findings, we conducted a robustness check after dropping dispatches with more than the top 2 percentile (>15) of fake remarks (Appendix A3.8). Finally, we conducted two other robustness tests (the tables have been omitted for brevity). Firstly, to ensure the effect is not driven by DA experience, we compared the effect for new and experienced DAs and found no significant difference. Lastly, we analyzed with an alternative dependent variable, $percent_success$ $(= \frac{success}{parcels})$, to address concerns regarding whether or not the effect could be workload-driven.

The robustness of our results is validated in all the above analyses detailed in Appendix 3.

## 9. Discussion and Concluding Remarks

The rise of e-commerce in the last two decades has presented immense growth opportunities for LMD firms like DHL and FedEx. Most e-commerce giants depend on LMD firms to manage deliveries, and these firms, in turn, rely on DAs for operational efficiency. DAs' conduct directly impacts the reputation and profitability of LMD firms. It is, therefore, essential to understand the factors that influence DAs' behavior and the consequences of their actions to minimize suboptimal outcomes. Our study highlights the significant impact of DA misconduct on operational performance. Collaborating with an Indian LMD firm and analyzing an extensive dataset using instrumental variable regression, we discover that an increase in one fake remark reduces the next day's successful deliveries by 1.60% and first-time-right deliveries by 1.86%. These results can be attributed to effort reallocation by DAs to conceal misconduct, triggering a cascading effect on productivity. We also provide several pieces of evidence to explain that fake remarks are unethical behavior shaped by broader factors and not a viable strategy for managing capacity. When workers strategically shirk on any given assignment, this also impacts their productivity on new tasks due to the time and effort involved in concealing their unethical behavior. Assuming a nominal commission for each successful delivery, our findings indicate the LMD firm faces million-dollar annual losses due to productivity spillovers resulting from fake remarks. Misconduct in delivery operations is not only a financial cost; it reduces worker productivity, increases operational costs, contributes to a higher carbon footprint, exacerbates customer attrition, and tarnishes partnerships with e-commerce entities. To address this, we advocate for designing and implementing mechanisms that preempt and address such behaviors.

[31] https://www.statalist.org/forums/forum/general-stata-discussion/general/1308457-endogeneity-issue-negative-binomial

Our analysis suggests several preventive and mitigation strategies for managers to alleviate the impact of fake remarks and design an efficient operational system. First, we advocate for open discussion of misconduct during end-of-day meetings, which can act as a deterrent. This approach will foster transparency and leverage peer dynamics and the fear of public reprimand, which, according to our analysis, prompts DAs to engage in less misconduct. Second, we recommend rotating DAs to new areas and refraining from assigning them COD parcels, which can disrupt patterns that facilitate misconduct. We also conclude that known levers to improve productivity, such as familiarity, might not work in the presence of misconduct. Third, delivery operations should minimize workload volatility and simplify delivery geography, reducing opportunities and motivations for DAs to engage in fake remark incidents. Finally, as a preventive strategy, we recommend designing predictive systems using machine learning by focusing on data and signals for the early detection of drivers who are at risk of entering fake remarks.

Even though fake remarks are a moral hazard, our context cannot be explained by the existing principal-agent literature because instrumental controls do not drive the DA behavior but are normative (refer to Bellos and Ferguson (2017) for details of instrumental vs. normative controls). The principal-agent theory assumes both the agent and the principal have a von Neumann-Morgenstern utility function (Ross 1973) and is used to guide a single decision-maker or a small group of decision-makers, where decision maker(s) "rationally" evaluate alternative decisions (Hauser and Urban 1979). In our case, however, we observe strategic interactions between multiple decision-makers exhibiting normative behavior rather than rational decisions. Few papers empirically quantify the impact of moral hazard, especially in settings where agents exhibit normative behavior. Further, we do so in a context (last mile delivery) where misconduct has not been studied in detail to the best of our knowledge. Additionally, we incorporate the unethical behavior and peer effects literature to establish the normative behavior of DAs and its repercussions in a context where information asymmetry is significantly high due to the spatial fragmentation of agents.

We believe our results are relevant across various compensation models and can provide insights for on-demand delivery services, such as groceries and courier mail, where drivers' effort reallocation and strategic decision-making play a critical role in shaping operational efficiency and service quality. Our research suggests that the phenomenon of fake remarks is a strategic, unethical behavior by drivers, which is not solely a function of the compensation structure but also a product of broader operational and behavioral factors. Furthermore, we believe that it is not possible to eliminate fake remarks in practice, meaning that appropriate measurement and mitigation strategies will always be relevant.

Several potential avenues for future research exist. Drawing parallels from Ibanez et al. (2018), who explored task completion sequence discretion in healthcare, a similar study in LMD could examine the influence of DAs' parcel delivery sequence autonomy on productivity. Additionally, considering the broader operational context, for instance, the nuanced relationship between DA productivity and the origin of parcels from different e-commerce platforms, delivery area characteristics, and the influence of DAs' perceptions of consignees (Altman et al. 2021), in understanding driver behavior could inform strategies to address and mitigate the challenges faced by drivers in different scenarios. We expect our findings to inspire more research into individual misconduct within service operations.